\documentclass[12pt]{article}
\usepackage[english]{babel}
\usepackage{mathtext}
\usepackage[T2A]{fontenc}
\usepackage{epsfig}

\begin{document}
\begin{center}
{\Large {\bf The Renyi entropy as a "free entropy" for complex systems. }}\\
\vspace{.23cm}
\large {\bf A. G. Bashkirov}\footnote{Institute for Dynamics of Geospheres RAS,
0119334, Moscow, Russia; e-mail: abas@idg.chph.ras.ru (A.G.Bashkirov)}\\

\end{center}
\begin{abstract}
The Boltzmann entropy $S^{(B)}$ is true in the case of equal
probability of all microstates of a system. In the opposite case
it should be averaged over all microstates that gives rise to the
Boltzmann--Shannon entropy (BSE). Maximum entropy principle (MEP)
for the BSE leads to the Gibbs canonical distribution that is
incompatible with power--low distributions typical for complex
system. This brings up the question:  Does the maximum of BSE
correspond to an equilibrium (or steady) state of the complex
system? Indeed, the equilibrium state of a thermodynamic system
which exchange heat with a thermostat corresponds to maximum of
Helmholtz free energy rather than to maximum of average energy,
that is internal energy $U$. Following derivation of Helmholtz
free energy the Renyi entropy is derived as a cumulant average of
the Boltzmann entropy for systems which exchange an entropy with
the thermostat. The application of MEP to the Renyi entropy gives
rise to the Renyi distribution for an isolated system. It is
investigated for a particular case of a power--law Hamiltonian.
Both Lagrange parameters, $\alpha$ and $\beta$ can be eliminated.
It is found that $\beta$ does not depend on a Renyi parameter $q$
and can be expressed in terms of an exponent $\kappa$ of the
power--law Hamiltonian and $U$. The Renyi entropy for the
resulting Renyi distribution reaches its maximal value at
$q=1/(1+\kappa)$ that can be considered as the most probable value
of $q$ when we have no additional information on behavior of the
stochastic process. The  Renyi distribution for such $q$ becomes a
power--law distribution with the exponent $-(\kappa +1)$.  Such a
picture corresponds to some observed phenomena in complex systems.

\end{abstract}
 \setcounter{footnote}{0}

\section{Introduction}
The well--known Boltzmann formula defines a statistical entropy,
as a logarithm of number of states $W$ attainable for the system
\begin{equation}
S^{(B)}=\ln W
\end{equation}
Here and below the entropy is written as dimensionless value
without the Boltzmann constant $k_B$.

This definition is valid not only for physical systems but for
much more wide class of social, biological, communication and
other systems described with the use of statistical approach. The
only but decisive restriction on the validity of this equation is
the condition that all $W$ states of the system have equal
probabilities (such systems are described in statistical physics
by a microcanonical ensemble). It means that probabilities  $p_i
=p_W=1/W$ (for all $i =1, 2, ..., W$) that permits to rewrite the
Boltzmann formula (1) as
\begin{equation}
S^{(B)}=- \ln p_W.
\end{equation}
When the probabilities $p_i$ are not equal we can introduce an
ensemble of microcanonical subsystems in such a manner that all
$W_i$ states of the $i$-th subsystem have equal probabilities
$p_i$ and its Boltzmann entropy is $S_i^{(B)}=- \ln p_i$. The
simple averaging of the Boltzmann entropy $S_i^{(B)}$ leads to the
Gibbs--Shannon entropy
\begin{equation}
S^{(G)}=\langle S_i^{(B)}\rangle_p \equiv - \sum_i p_i\ln p_i.
\end{equation}
Just such derivation of $S^{(G)}$ is used in some textbooks (see,
e. g. \cite{Haken,Nicolis}). This entropy is generally accepted in
statistical thermodynamics and conventional in a communication
theory.

According to maximum information principle (MEP) developed by
Jaynes \cite{Jaynes} for a Gibbs--Shannon statistics an
equilibrium distribution of probabilities $p=\{p_i\}$ must provide
maximum of the Gibbs--Shannon information entropy $S^{(B)}(p)=-k_B
\sum_i\,p_i\ln p_i$ upon additional conditions of normalization
$\sum_i p_i=1$ and a fixed average energy
\begin{equation}
U=\langle H\rangle_p\equiv\sum_i H_i p _i,
\end{equation}
where $H=\sum_i H_i$ is the Hamiltonian of the system.

Then, the distribution $\{p_i\}$ is determined from the extremum
of the functional
 \begin{equation}
L_G(p )=- \sum_i^W\,p_i\ln p_i - \alpha_0 \sum_i^W\,p _i - \beta_0
\,\sum_i^W H_i p _i ,
\end{equation}
where $ \alpha_0 $ and $\beta_0 $  are the Lagrange multipliers.
Its extremum is ensured by the Gibbs canonical distribution
\begin{equation}
p^{(G)}_i=Z_G^{-1}e^{- \beta_0 H_i},
\end{equation}
in which $\beta_0$ is determined by condition of correspondence
between Gibbs thermostatistics and classical thermodynamics as
$\beta_0 =1/k_BT_0$where $T_0$is the thermodynamic temperature.

However, when investigating complex physical systems (for example,
fractal and self-organizing structures, turbulence) and a variety
of social and biological systems, it appears that the Gibbs
distribution does not correspond to observable phenomena. In
particular, it is not compatible with a power-law distribution
that is typical \cite{Bak}  for such systems. Introducing of
additional restrictions to a sought distribution in the form of
conditions of true average values $\langle X^{(m)}\rangle_p$ of
some physical parameters of the system $X^{(m)}$ gives rise to a
generalized Gibbs distribution with additional terms in the
exponent (6) but does not change its exponential form.

\section{Helmholtz free energy and Renyi entropy}
There is no doubts about the MEP as itself, because of it is a
kind of "a maximum honesty principle" according to which we demand
a maximal uncertainty from the distribution apart from true
description of prescribed averages. In the opposite case we risk
to introduce a false information into the description of the
system.

Thus, it only remains for us to throw doubt on the information
entropy form. To seek out a direction of modification of the
Gibbs--Shannon entropy we consider first extremal properties of an
equilibrium state in thermodynamics.

A direct calculation of the average energy of a system gives the
internal energy (4), its extremum is characteristic of an
equilibrium state of rest for a mechanical system, other than a
thermodynamic system that can change heat with a thermostat. An
equilibrium state of the latter system is characterized by
extremum of the Helmholtz free energy $F$. To derive it
statistically from the Hamiltonian without use of thermodynamics
we introduce generating function \cite{Balescu}introduce
generating function for the random value $H =- \sum_i H_i$
\begin{equation}
\Phi_H (\lambda)=\sum_i e^{ \lambda H_i},
\end{equation}
where $\lambda$ is the arbitrary constant, and construct a
cumulant generating function
\begin{equation}
\Psi_H (\lambda)=\ln \Phi_H (\lambda)
\end{equation}
that becomes the Helmholtz free energy $F$ when devided by
$\lambda$ that is chosen as $\lambda  = -1/k_B T_0$.

Now we return to the problem of a generalized entropy for open
complex systems. Exchange by both energy and entropy is
characteristic for them. As an illustration, there is a
description by Kadomtsev \cite{Kadom} of self-organized structure
in a plasma sphere: "The entropy is being born continuously within
the sphere and flowing out into surroundings. If the  entropy flow
had been blocked, the plasma would 'die'. It is necessary to
remove continuously 'slag' of newly produced entropy".

That is the reason why the Gibbs--Shannon entropy, derived by the
simple averaging of the Boltzmann entropy can not be à function of
which extremum characterizes a steady state of a complex system
which exchange entropy with surroundings, just as the minimum of
the internal energy does not characterize an equilibrium state of
the thermodynamic system being in heat contact with a heat bath.

An effort may be made to find a "free entropy" of a sort by the
same way that was used above for derivation of the Helmholtz free
energy. The generating function is introduced as
\begin{equation}
\Phi_S (\lambda)=\sum_i e^{ \lambda S_i^{(B)}}
\end{equation}
Then the cumulant generating function is
\begin{equation}
\Psi_S (\lambda)=\ln \Phi_S (\lambda)=\ln\sum_i p_i^{- \lambda}.
\end{equation}
To obtain the desired generalization of the entropy we are to find
a $\lambda$-dependent numerical coefficient which ensures a
limiting pass of the new entropy into the Gibbs--Shannon entropy.
Such the coefficient is  $(\lambda+1)$. Indeed, the new
$\lambda$-family of entropies
\begin{equation}
S(\lambda)=\frac 1{1+\lambda}\ln\sum_i p_i^{- \lambda}.
\end{equation}
includes the Gibbs--Shannon entropy as a particular case when $
\lambda\to -1$.

Thus, it has appeared that the desired "free entropy" coincides
with the known Renyi entropy \cite{Renyi}. It is conventional to
write it with the parameter $q = -\lambda$ in the form
\begin{equation}
S^{(R)}(p)=\frac 1{1-q}\ln\sum_i p_i^{q}
\end{equation}

The same result can be obtained with the use of a
Kolmogorov--Nagumo \cite{Kolm,Nagumo} generalized averages of the
form
\begin{equation}
\langle x \rangle_\phi=\phi^{-1}\left(\sum_i p_i\phi (x_i)\right )
\end{equation}
where $\phi (x)$ is an arbitrary continuous and strictly monotonic
function and $\phi^{-1} (x)$ is the inverse function. It had been
just this kind of average that was used by Renyi \cite{Renyi2} to
define his new entropy as a generalized average of the Boltzmann
entropy (1). As a result of choice of the Kolmogorov--Nagumo
function in the form $\phi (x)=e^{(1-q)x}$ he obtained (12). Such
the choice of $\phi (x)$ appears accidental until it is not
pointed that the same exponential function of the Hamiltonian
provides derivation of the free energy that is extremal at an
equilibrium state of a thermodynamic system which exchange heat
with a heat bath. This fact permits to suppose that the Renyi
entropy derived in the same manner is extremal at a  steady state
of a complex system which exchange entropy with its surroundings
very actively.

Different properties of the Renyi entropy are discussed in
particular in Refs. \cite{Renyi,Beck,Klim}. It is positive
($S^{(R)}\geq 0$), convex at $q\leq 1$, passes into the
Gibbs--Shannon  entropy $\lim_{q\rightarrow 1}S^{(R)}=S^G$ and
into Boltzmann entropy (1) for any $q$ in the case of equally
probable distribution $p$.

In the case of $|1-\sum_i p_i^q|\ll 1$ (which, in view of
normalization of the distribution $\{p_i\}$, corresponds to the
condition $|1-q|\ll 1$), one can restrict oneself to the linear
term of logarithm expansion in the expression for $S^{(R)}(p)$
over this difference, and $S^{(R)}(p)$ changes to the Tsallis
entropy \cite{Tsallis}
\begin{equation}
S^{(T)}(p)=-\frac {k_B}{1-q}(1-\sum_i^W p_i^q).
\end{equation}
The logarithm linearization results in the entropy becoming
nonextensive, that is, $S(W_1W_2)\neq S(W_1)S(W_2)$. This property
is widely used by Tsallis and by the international scientific
school that has developed around him for the investigation of
diverse nonextensive systems (see web site \cite{Tsallis2}). In so
doing, the above-identified restriction $|1-q|\ll 1$ is
disregarded. As a result of nonextensivity the Tsallis entropy
incompatible with the Boltzmann entropy (1) because of the latter
was derived by Plank in such the form just from the extensivity
condition $S(W_1W_2)= S(W_1)S(W_2)$.

\section{MEP for Renyi entropy}
If the Renyi entropy is used in MEP instead of the Gibbs--Shannon
entropy, the equilibrium distribution is to be looked for from the
maximum of the functional
\begin{equation}
L_R(p )=\frac 1{1-q}\,\ln \sum_i^W\,p^{q } _i - \alpha \sum_i^W\,p
_i - \beta \,\sum_i^W H_i p _i,
\end{equation}
where $ \alpha $ and $\beta $  are Lagrange multipliers. It can be
noticed that $L_R(p)$ passes to $L_G(p )$ in the $q\to 1$ limit.

We equate a functional derivative of $L_R(p )$ to zero, then
\begin{equation}
\frac{\delta L_R(p )}{\delta p _i}=\frac q{1-q}\,\frac
{p_i^{q-1}}{\sum_j p_j^{q}} - \alpha -\beta H_i =0.
\end{equation}
To eliminate the parameter $\alpha$ we can multiply this equation
by $p_i$ and sum up over $i$, taking into account the
normalization condition $\sum_i p_i=1$. Then we get
\begin{equation}
\alpha =\frac q{1-q}- \beta   U
\end{equation}
and
\begin{equation}
p _i=\left((1-\beta\frac{q-1}{q} \Delta H_i)\,\sum_j^W\,p^{q} _j
\right)^{\frac{1}{q-1}},\,\,\Delta H_i=H_i-  U.
\end{equation}
Using once more the condition $\sum_i p_i=1$ we get $$
\sum_j^W\,p_j^q =\left(\sum_i^W (1-\beta\frac{q-1}{q}\Delta
H_i)^{\frac{1}{q-1}}\right)^{-(q-1)}$$ and, finally, we get
\cite{Bash1,Bash2} the Renyi distribution
\begin{eqnarray} p_i=p_i^{(R)}
&=&Z_R^{-1}\left(1-\beta\frac{q-1}{q}\Delta
H_i\right)^{\frac{1}{q-1}}\\
Z_R^{-1}&=&\sum_i\left(1-\beta\frac{q-1}{q}\Delta
H_i\right)^{\frac{1}{q-1}}.
\end{eqnarray}

When applying MEP to the Tsallis  entropy it is necessary to take
into account that all average values in the last version of
nonextensive thermostatistics \cite{Tsallis3} are calculated with
the use of an escort distribution $P_i=p_i^q/\sum_ip_i^q$  that
contradicts to the main principles of probability description.
Indeed, at $q>1$ ($q<1$), the importance of $p_i$ with the maximal
(minimal) values increases. In view of this, it is evident that
use of the escort distribution in statistical thermodynamics does
not lead to true average values of dynamical variables.
Nevertheless, the additional condition of a fixed mean energy in
nonextensive thermostatistics is written as $ U=\langle
H\rangle_{es}\equiv\sum_i H_i\,P_i$ and the escort Tsallis
distribution is found in the form
\begin{equation}
P_i^{(Ts)}=Z_{Ts}^{-1} \left(1-\beta^*(1-q')\Delta
H_i\right)^{\frac{q'}{1-q'}}\\
\end{equation}
where $\beta^*= \beta /\sum_ip_i^q$ and $\beta$ is the Lagrange
multiplier.

It should be noticed that both Renyi and escort Tsallis
distributions, Eqs. (19) and (21), are identical if $q'=1/q$. In
reality, in this case
\begin{equation}
1-q'=\frac{q-1}{q},\,\,\,\,\, \frac{q'}{1-q'}=\frac{1}{q-1}
\end{equation}
and $\beta^*$ is determined by the same second additional
condition (4) of MEP as well as $\beta$.

Thus, not always justified linearization of the logarithm in the
Renyi entropy and questionable use of the escort distribution
leads to the same Renyi distribution if $q$ and $q'$ are
considered as free parameters.

Therefore, numerous works (see  \cite{Tsallis2}) confirming
correspondence of the Tsallis escort distribution (with fitted
$q'$) with distributions in complex physical, biological, social
and other systems count rather in favor more justified Renyi
entropy than nonextensive Tsallis entropy.

When $q\to 1$ the distribution $\{p^{(R)}_i\}$ becomes the Gibbs
canonical distribution and $\beta/q\to\beta_0=1/k_BT_0$. Such
behavior is not enough for unique determination of $\beta$, as in
general, it may be an arbitrary function $\beta(q)$ which becomes
$\beta_0$ in the limit $q\to 1$.

To find an explicit form of $\beta$, we return to the additional
condition of the pre-assigned average energy and substitute there
the Renyi distribution (19). Then we obtain the integral equation
for $\beta$
\begin{equation}
U=\sum^W_i H_i p^{(R)}_i,
\end{equation}
where $U$ is considered as a known value. This equation was solved
\cite{Bash2} for a particular case of a power-law dependence of
the Hamiltonian on a parameter $x$
\begin{equation}
H_i=Cx_i^\kappa .
\end{equation}
This type of the Hamiltonian corresponds to an ideal gas model in
the Boltzmann-Gibbs thermostatistics and it seems reasonable to
say that it may be useful in construction of thermostatistics of
complex systems. Moreover, in most social, biological and
humanitarian sciences the system variable $x$ can be considered
(with $\kappa =1$) as a kind of the Hamiltonian (e.g. the size of
population of a country, effort of a word pronouncing and
understanding, bank capital, number of scientific publications of
an author, size of an animal etc.).

For the power-law Hamiltonian the convergence condition for the
sum in Eq. (23) in the limiting case $W\to\infty$ is
\begin{equation}
q>q_{min}=1/(1+\kappa)
\end{equation}
As a result of solution of equation (23) the parameter $\beta$ is
found as
\begin{equation}
\beta  =\frac 1{\kappa U}
\end{equation}
Independence of this relation from $q$ means that it is true, in
particular, for the limit case $q=1$ where the Gibbs distribution
takes a place and, therefore,
\begin{equation}
\beta=\beta_0\equiv 1/k_BT\,\,\,\,{\rm for \,\,all}\,\,q.
\end{equation}

When $H=p^2/2m$ (that is, $\kappa=2$) we get from (26) and (27)
that $U=\frac 1{2}k_BT_0$, as would be expected for
one-dimensional ideal gas.

The Lagrange parameter $\beta$ can be eliminated from the Renyi
distribution (19) with the use of Eq. (26) and we have,
alternatively,
\begin{equation}
p^{(R)}_i=Z^{-1}\left(1- \frac{q-1}{\kappa q}(C_u
x^\kappa_i-1)\right)^{\frac{1}{q-1}},\,\,\,H_i=Cx_i^\kappa,\,\,\,C_u=C/U.
\end{equation}
\begin{figure}[t]
\begin{minipage}{.50\linewidth}
 \centering\epsfig{figure=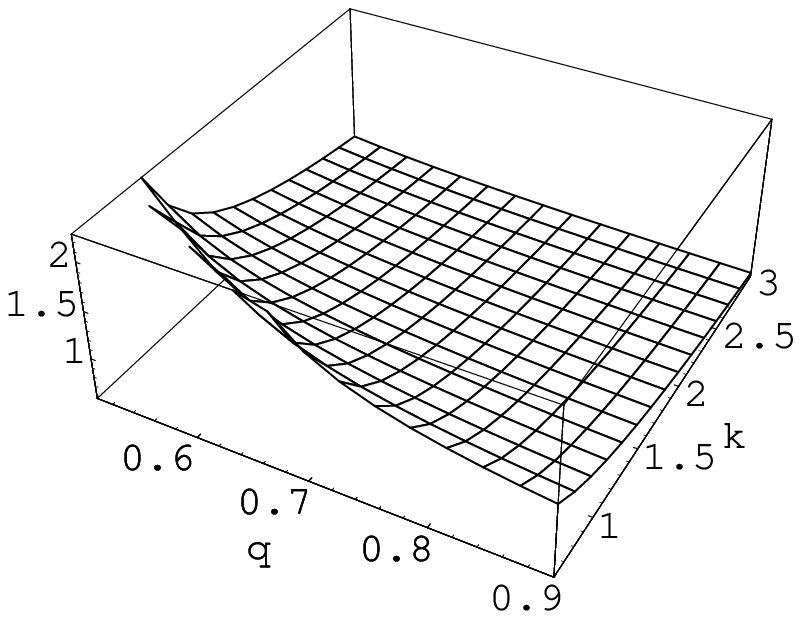, height=5cm}
 \end{minipage}
\begin{minipage}{.55\linewidth}
 \centering\epsfig{figure=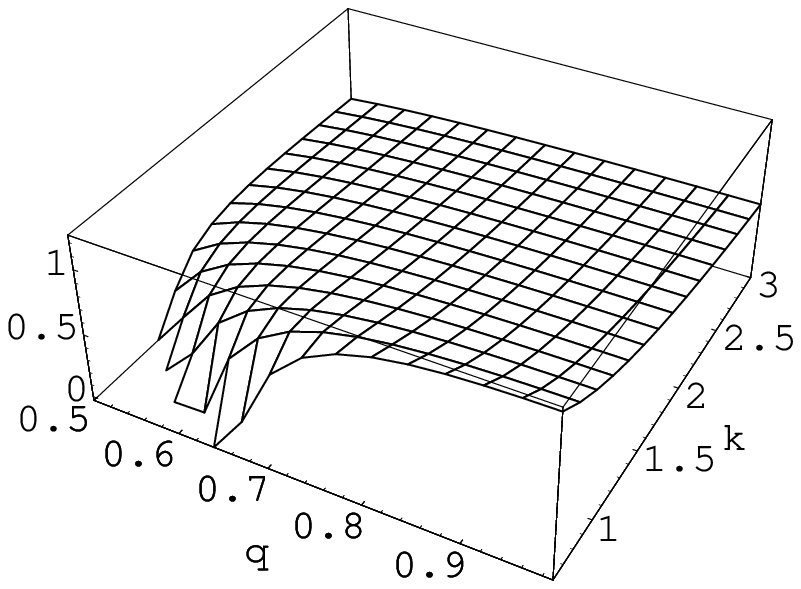, height=5cm}
  \end{minipage}
\caption{The entropies $S^{(R)}[p^{(R)}(q,\kappa)]$ (top) and
$S^{(B)}[p^{(R)}(q,\kappa)]$ (bottom) for the power--law
Hamiltonian with the exponent $\kappa$ within the range
$3>\kappa>0.5$ and $q>1/(1+\kappa)$.}
\end{figure}

The problem to be solved for a unique definition of the Renyi
distribution is determination of a value of the Renyi parameter
$q$. This will be the subject of the next section.

\section{The most probable value of the Renyi parameter}

An excellent example of solution of this problem for a physical
non-Gibbsian system was presented by Wilk and Wlodarczyk
\cite{Wilk}. They took into consideration fluctuations of both
energy and temperature of a minor part of a large equilibrium
system. This is a radical difference of their approach from the
traditional Gibbs method in which temperature is a constant value
characterizing the thermostat. As a result, their approach led
(see \cite{BaSuk}) to the Renyi distribution with the parameter
$q$ expressed via heat capacity $C_V$ of the minor subsystem
\begin{equation}
q=\frac {C_V-k_B}{C_V}.
\end{equation}
The approach by Wilk and Wlodarczyk was advanced by Beck
\cite{Beck1} and Beck and Cohen \cite{Beck2} who offered for it a
new apt term "superstatistics". In the frame of superstatistics,
the parameter $q$ is defined by physical properties of a system
which can exchange energy and heat with a thermostat. As a result,
$q\neq 1$ but $|q-1|\ll 1$, because of exchange entropy is not
taken into account by superstatistics.

In general, when we have no information about nature of a
stochastic process the parameter $q$ is considered as a free
parameter. So, the proposed further extension of MEP  consists in
looking for a maximum of the Renyi entropy in a space of the Renyi
distributions with different values of $q$.

In reality, maximum of the RE is ensured by the Renyi distribution
function (28) for any fixed $q$ fulfilling the inequality (25).
The next step consists of substitution of the Renyi distribution
$p^{(R)}(q,\kappa)$ into the definition of the Renyi entropy, Eq.
(13), and variation of the $q$-parameter. The resultant picture of
$S^{(R)}[p^{(R)}(q,\kappa)]$ as a function of $q$ is illustrated
in Fig. 1 (left). It is seen that $S^{(R)}[p^{(R)}(q,\kappa)]$
attains its maximum at the minimal possible value $q=q_{min}$. For
$q<q_{min}$, the series (23) diverges and, therefore, the Renyi
distribution does not determine the average value $U=\langle
H\rangle_p$, that is a violation of the second condition of MEP.

The similar procedure is applied to the Gibbs-Shannon entropy
$S^{(B)}(p)$, as well. Substituting there $p=p^{(R)}(q,\kappa)$ we
get the $q$-dependent function  $S^{(B)}[p^{(R)}(q,\kappa)]$
illustrated in Fig. 1 (right). As would be expected, the
Gibbs-Shannon entropy $S^{(B)}[p^{(R)}(q,\kappa)]$ attains its
maximum value at $q=1$ where $p^{(R)}(q,\kappa)$ becomes the Gibbs
canonical distribution.

Thus, it is found that the maximum of the Renyi entropy is
realized at $q=q_{min}$ and it is just the value of the Renyi
parameter that should be used if we have no additional information
on behavior of the stochastic process under consideration.

Recall that the Renyi entropy was derived above as a functional
which attain its maximum value at the equilibrium state (or steady
state) of a complex system just as the Helmholtz free energy for a
thermodynamic system. (The kind of extremum, that is, maximum or
minimum is determined by the sign definition for the constant
$\lambda$.) Then, a radically important conclusion follows from
comparision of these two graphs:

{\it In contrast to the Gibbs--Shannon entropy the Renyi entropy
increases as a system complexity (departure of the $q$ value from
1) increases that permits to explain an evolution of the system to
self-organization from the point of view of thermodynamical
stability.}

It counts in favor of such the conclusion that a power--law
distribution characteristic for self-organizing systems \cite{Bak}
is realized when the Renyi entropy is maximal. Substitution of
$q=q_{min}$ into Eq. (28) does lead to
\begin{equation}
p =Z^{-1} x^{-(1+\kappa)}
\end{equation}
Thus,  for $q=q_{min}$ the Renyi distribution for a system with
the power--law Hamiltonian becomes power--low distribution over
the whole range of $x$.

For a particular case of the impact fragmentation where $H\sim
m^{2/3}$ the power-law distribution of fragments over their masses
$m$ follows from (21) as $p(m)\sim m^{-5/3}$ that coincides with
results of our previous analysis \cite{fragm} and experimental
observations \cite{Fujiw}.

For another particular case, $\kappa =1$,  power--low distribution
is $p\sim x^{-2}$. Such a form of the Zipf-Pareto law is the most
useful in social, biological and humanitarian sciences. The same
exponent of power--low distribution was demonstrated \cite{Rybcz}
for energy spectra of particles from atmospheric cascades in
cosmic ray physics and for distribution of users among the web
sites \cite{Adamic}.

I acknowledge fruitful discussions of the subject with  A. V.
Vityazev.

\end{document}